\documentclass[prd, preprint, aps]{revtex4}

\usepackage{graphicx}
\usepackage{color}
\usepackage{amsfonts}

\def\be{\begin{equation}}
\def\ee{\end{equation}}
\def\ba{\begin{eqnarray}}
\def\ea{\end{eqnarray}}
\def\beq{\begin{eqnarray}}
\def\eeq{\end{eqnarray}  }

\def\rmd{{\rm d}}

\begin{document}

\newcommand{\myL}{\mathcal{L}}
\newcommand{\del}{\nabla}
\newcommand{\Real}{\mathbb{R}}
\newcommand{\interp}{I_{2h}^h\,}
\newcommand{\degree}{o}
\newcommand{\prolong}{I_{2h}^h\,}
\newcommand{\restrict}{I_{h}^{2h}\,}
\newcommand{\mkeq}[1]{\begin{equation}#1\end{equation}}

\title{The Volume Inside a Black Hole}

\author{Brandon DiNunno}
\author{Richard A. Matzner}

\affiliation{Center for Relativity, University of Texas at Austin,
Austin, TX 78712-1081, USA}

\begin{abstract} 
The horizon (the surface) of a black hole is a null surface, defined by 
those hypothetical ``outgoing" light rays that just hover under the 
influence of the strong gravity at the surface. Because the light rays are 
orthogonal to the spatial 2-dimensional surface at one instant of time, the 
surface of the black hole is the same for all observers ({\it i.e.} the same 
for all coordinate definitions of ``instant of time"). This value is $4 \pi 
(2Gm/c^2)^2$ for nonspinning black holes, with $G=$ Newton's constant, $c=$ 
speed of light, and $m=$ mass of the black hole.

The 3-dimensional spatial volume inside a black hole, in contrast, depends 
explicitly on the definition of time, and can even be time dependent, or 
zero. We give examples of the volume found inside a standard, nonspinning 
spherical black hole, for several different standard time-coordinate 
definitions.

Elucidating these results for the volume provides a new pedagogical resource
of facts already known in principle to the relativity community, 
but rarely worked out.

\end{abstract}

\pacs{ }

\maketitle

\vspace{0.5cm}


%
%
%
\section{Introduction}
\label{sec:intro}
The area of the surface of a nonspinning, quiescent (Schwarzschild) black hole 
is is $16 \pi (Gm/c^2)^2$, which is written $16 \pi 
m^2$ in the usual relativist's convention in which units are chosen so that 
both Newton's constant and the speed of light are set equal to unity. The 
usual Schwarzschild coordinate $r$ is defined to be an areal 
coordinate (spherical area = $4 \pi r^2$), so 
its value at the black hole surface (the {\it horizon}) is $r=2m$. Because 
the black hole is spherical, we simply need to measure the area in a 
transverse direction. This produces the unique result ($Area = 16 \pi 
m^2$). The uniqueness follows because if we consider a different 
definition of the 3-space in which we measure the area, we just shift 
our points in null directions along the (null) horizon.  Null directions 
have zero length and cannot contribute to (or change) the area. 

An occasional question to the teacher of relativity is : ``...then, what is 
the {\it volume} of a black hole?" The answer is that, unlike the response 
about the surface, the volume depends on the way that the 3-dimensional 
``constant-time" space containing the black hole is defined. Gravity, 
described by General Relativity, is the curvature of space-time, and the 
implied curvature for the $\it space$ defined by our choice of constant 
time depends on how the ``now" space is defined. 

The simplest black hole is an {\it eternal} black hole, one that was not 
formed by collapsing matter but is a nonlinear ``vacuum" solution, with 
structure anchored in its own gravitational field.  Even in this case there 
are many choices of constant time, and hence many different results for the 
volume, of the chosen 3-space within the horizon. This article presents a 
pedagogical description of that (well-known to the expert) fact.

All of the background needed for this paper can be found 
in {\it Gravitation} \cite{MTW}, henceforth {\it MTW}. 
We will also provide original references where appropriate.

\section{Background} 
\label{sec:Background}

Einstein's General Relativity describes the gravitational field by giving 
the spacetime metric. A metric describes the way coordinate increments apply 
to measurable (``proper") space or time increments. The {\it Special} 
Relativity metric (describing a spacetime without gravity) is written:

\be 
\rmd s^2 = -\rmd t^2 +  \delta_{ij}\, \rmd x^i\, \rmd x^j 
\label{eq:flatMetric} 
\ee

Here we use the {\it summation convention}: repeated indices are summed over 
through their range; variables $i, j, k, ...$ range through the spatial 
coordinates $x, y, z$, and $\delta_{ij} =1$ if $i=j$, and $\delta_{ij} =0$ 
if $i \ne j$. (the superscripts $i, j, k, ...$ are indices, not exponents.) 
In Eq.(\ref{eq:flatMetric}), $s$ is a proper distance, so to 
make the units match, the first term on the right should have a factor 
$c^2$, the square of the speed of light. As noted above, relativists 
simplify expressions by using units in which the speed of light is unity. 
(For instance: the length unit is the light year and the time unit is the 
year.)  Also, it is useful to rewrite expression (1) using spherical 
coordinates:

\be 
\rmd s^2 = -\rmd t^2 +  f_{ij}\, \rmd x^i\, \rmd x^j 
\label{eq:flatMetricSpherical} 
\ee
where now $i, j, k, ...$ range through the spatial coordinates $r, \theta, 
\phi$, and $f_{ij}$ is  a diagonal symmetric matrix $f_{ij} = diag[1, r^2, 
r^2 sin^2\theta] = diag[1, (x^1)^2, (x^1)^2 sin^2 (x^2)]$. The metric is an 
example of a tensor, a geometrical object that is defined independently of 
any particular coordinate frame, and whose components follow specific rules 
for expression in different reference frames. Both Eq(\ref{eq:flatMetric}) 
and Eq(\ref{eq:flatMetricSpherical}) present the same geometrical object, 
the Special Relativity metric tensor, but expressed in different coordinate frames.

Within a year of the publication of Einstein's General Relativity \cite{eins}, 
Schwarzschild \cite{schw} obtained the General Relativity metric which is the analog to 
the simplest Newtonian gravitational field with $\Phi = G 
m/r$.  (Don't confuse the coordinate $\phi$ with the potential $\Phi$.) 
This generalizes Eq.(\ref{eq:flatMetricSpherical}) to

\be 
\rmd s^2 = -(1 - \frac{2 \Phi}{c^2})\rmd t^2 +  (1 - \frac{2 
\Phi}{c^2})^{-1} \rmd r^2 +r^2 \rmd \theta^2 + +r^2 sin^2 \theta \rmd 
\phi^2. 
\label{eq:Schwarschild} 
\ee

For large distances from the center, the Schwarzschild form 
Eq(\ref{eq:Schwarschild}) indicates a moderate deviation from the Special 
Relativity form, and it can be shown that geodesic motion in this spacetime 
approximates Newtonian motion quite closely. However, if one considers 
smaller radii, the quantity

\be 
\frac{2\Phi}{c^2}=\frac{2Gm}{c^2 r}\equiv \frac{2m}{r}
\label{eq:potential} 
\ee
can become significant, and apparently cause problems as it approaches unity 
(the coefficient of $\rmd r^2$ diverges to infinity; the 
coefficient of $\rmd t^2$ goes to zero). This situation was confusing 
because objects with $\frac{2Gm}{r} \approx 1$ were obviously extremely compact, so 
maybe this was a nonphysical configuration; but orbits, particularly 
expressed in terms of the proper time of the infalling observer, showed this 
strange surface could be reached in the finite lifetime of the intrepid 
explorer willing to fall into the center of the field. 
Only in 1960 did Kruskal \cite{krus} and Szekeres \cite{szek} recognize that the surface 
$r=2m$ {\it is} special but is not singular, and show how to understand this 
fact.  (Here and henceforth we again set both $c$ and the Newtonian 
constant $G$ equal to unity.)

The trick is to realize that the Schwarzschild coordinates are badly behaved 
near $r=2m$, and to introduce new time and radial coordinates (called $v$, 
the time coordinate, and $u$, the new radial coordinate) which behave well 
(``smoothly") there. (The angles $\theta, \phi$ just describe 
2-dimensional spheres, and there is no reason to change {\it them}.) To make the
new coordinates behave well near $r=2m$ 
requires {\it singular} transformations \{$t,r$\}$\leftrightarrow$\{$u,v$\}, 
but this is justified by the fact 
that all particle and photon orbits remain smooth and continuous when 
expressed in the new coordinates. Even more usefully, $u$ and $v$ are 
defined so that the radial coordinate speed of light , $\frac{du}{dv}=\pm 1$ 
everywhere. The null lines are inclined at $45^\circ$ just as they are in a
flat space diagram. This makes it easy to pick out timelike motion, or 3-spaces of 
constant time.

The transformations giving \{$t,r$\} in terms of \{$u,v$\} are:

\be 
(\frac{r}{2m} -1) e^{\frac{r}{2m}} = u^2-v^2.
\label{eq:r(u,v)} 
\ee

\be 
tanh ~\frac{t}{4m} = \frac{v}{u},~~~~~ |\frac{v}{u}| \le 1
\label{eq:t(u,v)} 
\ee

\be 
tanh ~\frac{t}{4m} = \frac{u}{v},~~~~~~ |\frac{v}{u}| \ge 1
\label{eq:t2(u,v)} 
\ee

\medskip

In Eq(\ref{eq:r(u,v)}) there is no analytic inverse for the single valued 
function of $r$ on the left hand side, but numerical solution is straightforward.
The fact that there are two different analytical expressions relating $t$ to \{$u,v$\}
is partial evidence of the singular coordinate transformation. 
The inverse transformation giving the Kruskal-Szekeres coordinates \{$u,v$\} in terms of \{$r, t$\}, 
can be found in {\it MTW}. 

What is more useful than the analytic expressions Eqs(\ref{eq:r(u,v)} - \ref{eq:t2(u,v)})
is to graph the lines showing constant Schwarzschild coordinates $t$ and 
$r$, in a graph whose axes are the  Kruskal-Szekeres coordinates $u$ and 
$v$. See Figure 1.
\begin{figure}
\begin{center}
\includegraphics[width=6.0in,angle=0]{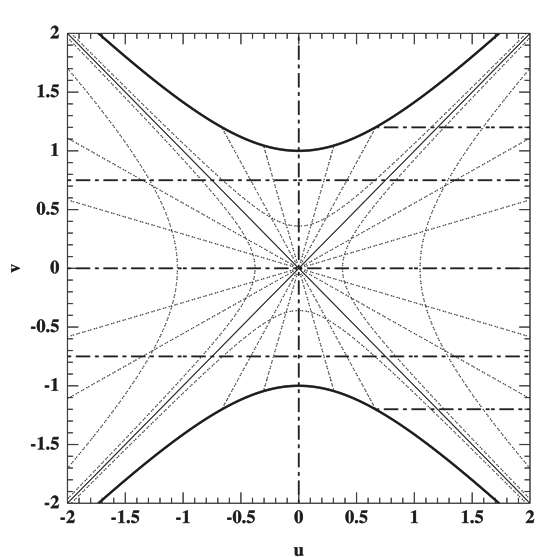}
\end{center}
\caption{The lines where the Schwarzschild coordinate $t=constant$, 
plotted in the Kruskal $\{$u,v\} plane (straight ``double-dot" lines 
passing through the origin); and the hyperbolae 
where the Schwarzschild radial coordinate $r=constant$ (``double-dot" curves). 
We also show several values of Kruskal coordinate $v=constant$ 
(heavy ``dot-dash" horizontal lines. The spacelike hyperbolae $v^2-u^2 = 1$ 
(heavy curves) are the locations where the Schwarzschild $r=0$. 
The curvature tensor is singular at $r=0$.  }
\label{fig:UV}
\end{figure}

From Figure 1 we see that constant Schwarschild coordinate $t$ is a straight line 
through the $u,v$ origin; $t = 0$ is the horizontal line coinciding with the 
line $v=0$, $t= \infty $ is the $45^\circ$ line $v=u$ (light solid positive-slope line in Figure 1 passing through the origin),  $t= - \infty $ 
is the $45^\circ$ line $v=-u$ (light solid negative-slope line in Figure 1 passing through the origin).
From Eq(\ref{eq:r(u,v)}), constant Schwarzschild coordinate $r$ defines a hyperbola 
given by $u^2-v^2$ = $const$. If the constant is positive, this defines the 
situation far from the black hole. However, if one considers smaller values 
for this constant, corresponding to smaller constant values of $r$, 
the hyperboloids approach the
straight lines $u=\pm v$, which are achieved when $r=2m$, when the left hand side 
of Eq(\ref{eq:r(u,v)}) is zero. Thus $r=2m$ overlays (defines 
the same points as) $t = \pm \infty$. Outside the horizon, $r=const$ is a 
timelike surface (its tangent is a timelike vector); inside the horizon it is spacelike.

\section{Volume Computation: Schwarzschild Coordinates} 
\label{sec:SchVol}

The volume is a three dimensional concept, and it depends only on the $t=constant$ spatial 
3-dimensional part of the metric. This means, consider the 4-d metric (Eq (3)) with  
the differential of the time coordinate set equal to zero ($dt=0$ in the 
Schwarschild coordinates we consider in this section). From the resulting 
3-dimensional metric, compute the determinant, $g$, of the matrix of metric components.

\be 
g = \frac{r^4 sin^2 \theta}{1 - \frac{2m}{r}}, ~~~~~~ Schwarzschild~coordinates.
\label{eq:schwDet} 
\ee

The volume between two values of $r$, say $r_{inner}$ and $r_{outer}$ is then

\be 
\int^{r_{outer}}_{r_{inner}} \sqrt{g} d^3x.
\label{eq:schwVol}
\ee

We have been asked to compute the volume inside the horizon at a fixed 
Schwarzschild time $t$. Thus the 
outer limit in the integral in Eq(\ref{eq:schwVol}) is $r_{outer}= 2m$.

Looking at Figure (1), we see that on {\it no} 
Schwarzschild 
$t=constant$ ``slice" does the $r$ coordinate extend to 
less than $r=2m$. 
The limits are the same: $r_{inner}=r_{outer}=2m$, so we expect the integral to vanish. 
However, the integrand is singular at $r=2m$, so we must investigate the integral a 
little further.

The angular integral just yields $4\pi$. The integral in $r$ has a singularity at $r=2m$, 
but this singularity is in fact integrable. We can investigate this by assuming the 
value of $r_{outer}$ to be just larger than $2m$, say $r_{outer}= 2m(1+\epsilon)$, where 
$\epsilon $ is small. Then to lowest order in $\epsilon$, the radial integral is 

\be 
\int^{r_{outer}}_{r_{inner}} \frac{(2m)^\frac{5}{2}dr}{\sqrt{r-2m}} 
\label{eq:schwEstimate}
\ee

\be 
\approx 2 \sqrt{r-2m} (2m)^\frac{5}{2}
\label{eq:schwEstAnswer}
\ee
evaluated at the limits. This integral obviously vanishes 
as $\epsilon$ goes to zero, i.e. as 
$r_{outer} \rightarrow 2m$.
{\it There is zero volume inside the black hole in any Schwarzschild time 
slice of a Schwarzschild black hole spacetime. }

\section{Volume Computation: Kerr Schild Coordinates} 
\label{sec:KSVol} 

Convincing oneself of the zero result of Section 3 is made easier 
by considering a different time-independent form of the 4-dimensional metric 
describing the Schwarzschild spacetime, which gives a nonzero volume inside the horizon.
Kerr-Schild coordinates provide this form.

Coordinates called Kerr-Schild coordinates \cite{KS}, which we denote $r_{KS}, t_{KS}$,
are related (in our spherical case) to the Schwarzschild coordinates by the transformations:

\be 
r_{KS} = r
\label{eq:rKS(r)}
\ee

\be 
t_{KS} = t + 2m ~ ln (\frac{r}{2m} -1)
\label{eq:tKS(r)}
\ee

The form of the 4-dimensional metric in Kerr-Schild coordinates is:

\be 
\rmd s^2 = -(1 - \frac{2m}{r_{KS}})\rmd t_{KS}^2 +   \frac{4m}{r_{KS}}\rmd t_{KS} \rmd r_{KS} 
+(1 + \frac{2m}{r_{KS}})\rmd r^2 +r^2 \rmd \theta^2 + r^2 sin^2 \theta \rmd 
\phi^2. 
\label{eq:KSmetric} 
\ee

The Kerr-Schild form of the metric is, like the Schwarschild form, independent of 
the time coordinate (here $t_{KS}$). It does however contain terms that describe 
cross terms in the 
measurement of distance, cross terms between increments in time and increments in 
radial coordinate. For the 3-d metric, however, these terms are irrelevant. The 
3-metric is simply 

 \be 
{}^3 \rmd s^2_{KS} = (1 + \frac{2m}{r_{KS}})\rmd r^2 +r^2 \rmd \theta^2 
+r^2 sin^2 \theta \rmd 
\phi^2
\label{eq:KSmetric3d}.
\ee

It is also important to know where the points in the constant $t_{KS} = constant $
slice are located with respect to the standard Kruskal-Szekeres picture. Figure 2 
shows a series of $t_{KS} =constant$ 
surfaces. 
\begin{figure}
\begin{center}
\includegraphics[width=6.0in,angle=0]{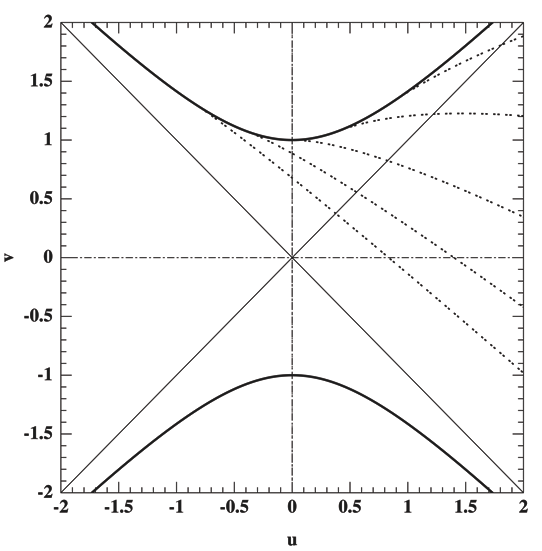}
\end{center}
\caption{The curves (``double-dot" curves) where the Kerr-Schild coordinate 
$t_{KS}=constant$, plotted in the Kruskal $\{$u,v\} plane. The radial 
coordinate in the Kerr-Schild system is the same as the standard Schwarzschild coordinate.}
\label{fig:TKSC}
\end{figure}
It can be seen that these constant $t_{KS}$ slices differ significantly 
from the Schwarzschild slices. They penetrate inside $r=r_{KS}=2m$ 
(which the constant Schwarzschild slices do not), and in fact extend 
inward to $r=r_{KS}=0$. We thus expect a nonzero result from  calculation of 
the volume inside the sphere $r=r_{KS}=2m$.

The determinant of the 3-metric Eq(\ref{eq:KSmetric3d}) is 
$(1 + \frac{2m}{r_{KS}})r_{KS}^4sin^2 \theta$. Thus the volume at constant $t_{KS}$ is

\be 
\int^{2m}_0 \sqrt{(1 + \frac{2m}{r_{KS}}) ~ r_{KS}^4 ~ sin^2 \theta} ~ dr_{KS} d\theta d\phi.
\label{eq:KSVol}
\ee
\be 
~~~~ = \int^{2m}_0 \sqrt{(1 + \frac{2m}{r_{KS}})} ~ r_{KS}^2 ~ sin \theta dr_{KS} d\theta d\phi.
\label{eq:KSVol2}
\ee

Even though the integrand contains the first factor which becomes infinite 
at $r_{KS} =0$ as $1/\sqrt{r_{KS}}$, the factor $r_{KS}^2$ moderates the 
divergence, and the integral is finite. In fact, since we consider the volume 
for $r_{KS} \le 2m$, after doing the angular integrals the integrand for the $r_{KS}$ 
integration is clearly between $4 \pi \sqrt{2m}(r_{KS})^\frac{3}{2}$
and $\sqrt{2}\times 4 \pi \sqrt{2m}(r_{KS})^\frac{3}{2}$ so the volume is easily analytically 
estimated to  be  between $\frac{2}{5} 4 \pi (2m)^3 = (5.026...)\times (2m)^3$, and  
$\sqrt{2} \times \frac{2}{5} 4 \pi (2m)^3 = (7.108...)\times (2m)^3$. 
The complete integral can be done analytically 
(Mathematica helps), yielding:

\be 
 Vol_{KS} =\frac{1}{24} [7\sqrt{2}- \frac{3}{2} ln (\frac{\sqrt{2}-1}{\sqrt{2}+1})]4 \pi (2m)^3.
\label{eq:KSVolEvaluated}
\ee
\be 
~~~~~~~ = ~ (6.567...) \times (2m)^3.
\nonumber
\ee

\section{Volume Computation: Novikov Coordinates} 
\label{sec:Novikov}

Both the Schwarschild and the Kerr Schild definitions of time  are such that 
the metric (and thus the volume inside the horizon) is static, {\it i.e.} 
independent of the Schwarzschild time $t$, or of the Kerr-Schild time $t_{KS}$ 
at which it is computed. However it is 
easy to see that a collection of nearby observers falling into a black hole 
would recognize a time dependent gravitational field; the tidal force gets stronger as they fall in, getting older. Novikov \cite{novi} realized that 
one could use the {\it proper} time $\tau$, the local internal time of each 
observer, to define a time slicing. He based his coordinates on a collection of
observers who at one Schwarzschild instant (Schwarzschild $t=0$, also labeled $\tau = 0$) 
are instantaneously each at rest at their own maximum value $r_{max}$ of the Schwarzschild 
coordinate $r$. 
Each observer sets her watch to read zero at this one instant of time.
Novikov  also introduced a radial coordinate, called a
{\it comoving} coordinate $R$, defined by: for each observer, 
\be 
 R^2+1= \frac {r_{max}}{2m},
\label{eq:NovikovCoorDef}
\ee
At any later Novikov time $\tau$, each
particle still has the same value of $R$ but has fallen to a smaller value of $r$. 
For general values of $\tau \ne 0$ one  
has only an implicit 
functional relation between $r$ and $R$: 

\be 
\frac{\tau}{2m} =  (R^2+1)[\frac {r}{2m}-\frac {({r/2m})^2}{R^2+1}]^{1/2} +(R^2+1)^{3/2} 
arccos [(\frac{r/2m}{R^2+1})^{1/2}].
\label{eq:ImplicitTau}
\ee

For each value of $\tau$ this is an implicit function giving $r(R)$, and $R(r)$ in that 
particular $\tau=const$ Novikov 3-space. Eq(\ref{eq:ImplicitTau}) holds for positive $\tau$. 
For negative $\tau$, use the fact that the relation between $r$ and $R$ is even in $\tau$:
$r(R, \tau)= r(R, -\tau)$, and  $R(r, \tau)= R(r, -\tau)$.
Once we have determined this relation, we can (by taking the differential of 
Eq (\ref{eq:ImplicitTau}) and setting $d \tau=0$) evaluate 
$\partial r/\partial R$ and $\partial R/\partial r$ at any time $\tau$. 
Figure 3 shows the relation between Novikov and Kruskal-Szekeres coordinates. 
\begin{figure}
\begin{center}
\includegraphics[width=6.0in,angle=0]{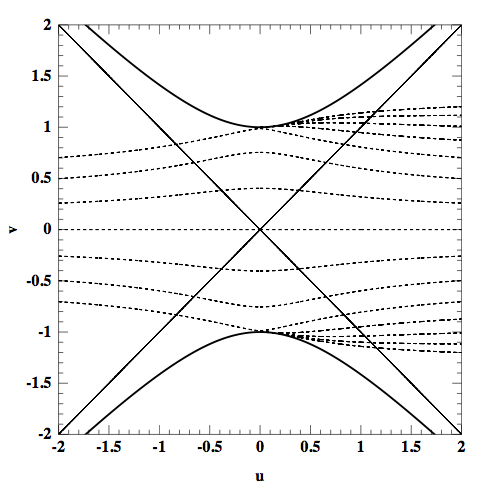}
\end{center}
\caption{The surfaces $\tau=constant$ (curved, roughly horizontal lines) plotted 
in the Kruskal diagram. $\tau=0$ coincides with $v=0$, and the figure is reflection 
symmetric across $v=0$. $\tau$ is the proper time of infalling observers, and is 
used as the time coordinate in Novikov coordinates. }
\label{fig:NK}
\end{figure}

The Novikov coordinate 
4-metric is: 

\be 
\rmd s^2 = -\rmd \tau^2    
+( \frac{R^2+1}{R^2})(\frac{\partial  r}{\partial R})^2 ~  \rmd R^2 +r^2 
(\rmd \theta^2 + sin^2 \theta \rmd \phi^2). 
\label{eq:NovMetric} 
\ee
The appearances of $r$ in this equation should be eliminated so that only 
$R$ and $\tau$ appear (using Eq(\ref{eq:ImplicitTau})), 
but we keep the compact symbol $r$ to mean the function $r(R, \tau)$.

The determinant of the 3-metric is
\be 
( \frac{R^2+1}{R^2})(\frac{\partial  r}{\partial R})^2 ~ r^4 sin^2 \theta.
\label{eq:NovDet} 
\ee

Thus the volume to be evaluated is 
\be 
\int^{R_{outer}}_{R_{inner}} \sqrt{( \frac{R^2+1}{R^2})} ~ |\frac{\partial  r}{\partial R}| 
r^2 sin \theta ~ dR ~ d \theta ~ d \phi
\label{eq:NovVol} 
\ee

The integrand is time dependent, because the factor $|\partial r/\partial R|$
depends on the Novikov time $\tau$. But, additionally, the {\it limits} of the 
integral are time dependent. For instance larger and larger values of $R$ fall 
toward the horizon as $\tau$ increases. Hence the value $R_{outer}$ increases monotonically 
as $\tau$ increases. Because we have access to the factor 
$(\partial r/\partial R)$, we can freely transform the integral between 
one expressed in coordinate $R$, and one in coordinate $r$. The 3-space $\tau = 0$
is identical to the Schwarzschild 3-space $t=0$, so the volume evaluated 
for $\tau =0 $ is the same as found for the standard Schwarschild coordinate case: $Vol=0$. 
This is consistent with the result from the integral (Eq(\ref{eq:NovVol})) evaluated in terms of Schwarschild 
coordinate $r$. The upper limit in every case is $r_{outer}=2m$. The lower limit 
on the initial ($\tau =0$) space is $r_{inner}=2m$ also, because this is the minimum 
$r$ reached in the 3-space $\tau=0$. At later $\tau$, since we are looking for the volume 
contained inside the horizon, $r_{outer}=2m$ continues to hold.
However, $r_{inner}$ decreases, because the observer originally at $r=2m$ 
({\it i.e.} at $R=0$) falls inward. 
That person falls inward for a time of $\pi m$, whereupon she reaches $r=0$ and her 
world line terminates (she is destroyed) because of the arbitrarily large tidal 
forces at $r=0$. She is the first observer to reach $r=0$. The $R$ for each infalling 
observer stays fixed at its 
initial value, and in particular the first observer's
Novikov coordinate stays at $R=0$.

Thus, between $\tau = 0$ and $\tau = \pi m$, the inner boundary for the integral 
is $r(R=0,\tau)$, a value of $r$ that has to be evaluated numerically, but which 
decreases monotonically from $r=2m$ to zero. We thus expect the volume inside the 
horizon to increase during this time. 

Note also (most easily seen from the diagram) that 
$r_{inner} $ is the smallest value of $r $ reached, but that the space 
inside the horizon on the left contains an equal volume, and is connected 
on this 3-space at $r_{inner}$. Hence between $\tau = 0$ and $\tau = \pi m$ the 
correct integration contains a factor of $2$: $2 \times \int^{2m}_{r_{inner}}$.

For $\tau > \pi m$ the singularity at $r=0$ intrudes to reduce the volume to that 
on only one side of $R=0$. Hence there is an instantaneous drop by half in the 
volume, at the time $\tau = \pi m$.

Subsequently to $\tau = \pi m$ the limits in terms of $r$ remain fixed at $0$, $2m$.
However the volume is still a function of time because the quantities involving 
R in the integrand have different dependences on r, for different values of $\tau$.

The resulting $\tau$- dependent volume is plotted in Figure 4.
\begin{figure}
\begin{center}
\includegraphics[width=5.5in,angle=0]{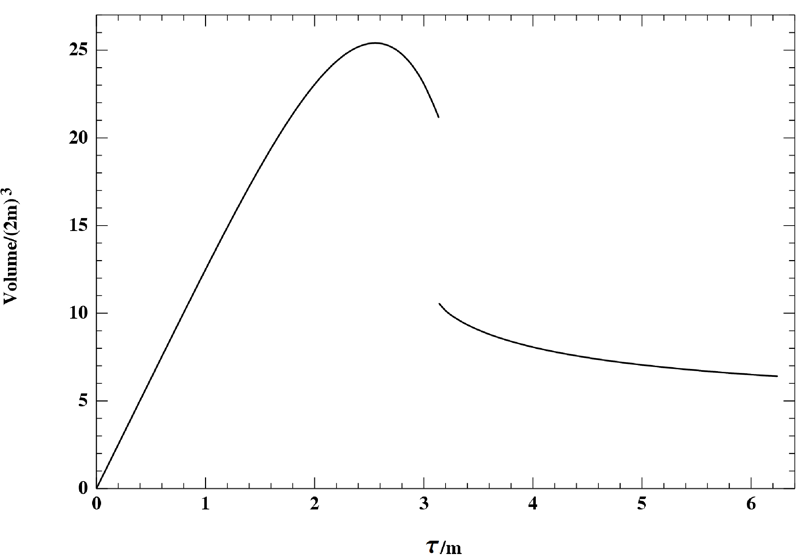}
\end{center}
\caption{The volume inside the horizon in Novikov coordinate 3-spaces, as a function of the Novikov time $\tau$, which is also the proper time of the infalling observers defining the coordinates. }
\label{fig:novVol}
\end{figure}

\section{Volume Computation: Kruskal-Szekeres Coordinates} 
\label{sec:Kruskal}

The 4-metric for the Scharzschild spacetime expressed in Kruskal-Szekeres coordinates is:

\be 
\rmd s^2_K = \frac{32m^3}{r} e^{-\frac{r}{2m}} (-\rmd v^2    + \rmd u^2) 
+r^2 (\rmd \theta^2 + sin^2 \theta \rmd \phi^2). 
\label{eq:KruskalMetric} 
\ee
where one must view the appearance of the function $r$ as shorthand for the function 
$r(u,v)$ defined by Eq(\ref{eq:r(u,v)}). We can investigate the time dependent ({\it i.e.}
$v${\it -dependent}) volume inside the horizon in a $v=const$ 3-space
in close analogy to the Novikov analysis above.

In fact, Figure 1 shows that the 3-spaces $v=const$ of the Kruskal-Szekeres coordinates 
are like straightened versions of the Novikov 3-spaces $\tau = const$. 
Completely analogously to the Novikov case, an 
observer initially at $r=2m$, {\it i.e.} $u=0$, falls inward for a time of $v=1$, whereupon 
she reaches $r=0$ and her 
world line terminates (she is destroyed) because of the arbitrarily large tidal 
forces at $r=0$. She stays at the
Kruskal-Szekeres coordinate $u=0$ as she falls.

Rather than compute by using the transformation to the coordinate r as we did for 
the Novikov case, we find it simpler to compute the volume directly in the Kruskal-Szekeres 
coordinates, where the limits on the radial integration are $u_{inner}=0$ and 
$u_{outer} = v$ so long as 
$v\le 1$, and $u_{inner}=\sqrt{v^2-1}$, $u_{outer} = v$ when $v>1$.

The determinant of the $v=const$ ~ 3-metric is
\be 
(\frac{32m^3}{r} e^{-\frac{r}{2m}}) r^4 sin^2 \theta. 
\label{eq:KruskalDet} 
\ee
Thus the volume to be evaluated is:

\be 
\int ^{u_{outer}}_{u_{inner}} \sqrt{\frac{32m^3}{r} e^{-\frac{r}{2m}}} ~ r^2 sin\theta ~ du ~ d\theta ~ d\phi. 
\label{eq:KruskalInt} 
\ee
\begin{figure}
\begin{center}
\includegraphics[width=5.5in,angle=0]{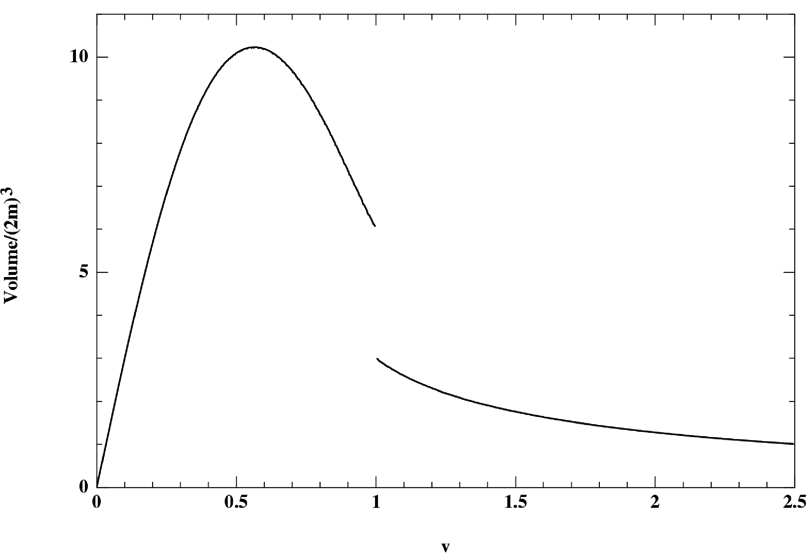}
\end{center}
\caption{The volume inside the horizon on a constant Kruskal-Szekeres time 3-space $v= constant$ 
as a function of $v$. The factor-of-two drop at $v=1$ arises because ``the throat pinches off" 
at that instant, separating two halves of the volume inside the horizon. }
\label{fig:kruskalVolume}
\end{figure}

Again as before, the initial ($v=0$) slice corresponds to the Schwarzschild coordinate 
3-space, so the volume is zero. (This can also be seen because it is an integration with
finite integrand
from $u=0$ to $u=v$ when $v$ is zero.) For somewhat larger $v$, the limits expand,  
and the integral is nonzero. For spaces before the singularity appears, we include a 
factor $2$ because the space extends to the horizon ``on the left"; at the time $v=1$ 
the singularity appears in the space (the 3-space touches the singularity; the 3-space 
evolves a singularity) and the volume drops by half. Subsequently the evolution is over 
limits $\sqrt{v^2-1}$, $v$, and the volume continues to evolve. Figure 5 presents 
the time evolution of the volume.

\section{Conclusions} 
\label{sec:Conclusions}

The area of a Schwarzschild black hole is unique, and can be defined by an idealized 
transverse measurement of a particular spherical surface. In contrast, the volume inside 
a black hole requires a definition of the particular 3-space in which the volume is computed, 
which may be explicitly time dependent, and an understanding of the (possibly time dependent) 
limits of the integral required to compute this volume. Understanding these points and 
computing the volumes as we have done here introduces and uses a number of concepts and
techniques of general use in relativistic calculation, and can be a useful pedagogical tool.

\section*{Acknowledgments}
This work was supported by NSF grant
PHY~0354842, and by NASA grant NNG04GL37G.  We thank Michael Salamon for bringing this 
problem to our attention.

\end{document}